\documentclass[11pt]{article}
\usepackage{amsmath,amsfonts,latexsym,graphicx,amssymb}
\usepackage[dvips]{epsfig}
\usepackage{amsfonts}
\usepackage{bm}
\date{}

\newcommand{\ot}{{\,\otimes\,}}
\newcommand{{\Cd}}{{\mathbb{C}^d}}

\newcommand{\sbalpha}{{\mbox{\scriptsize \boldmath $\alpha$}}}

\newcommand{\balpha}{{\mbox{ \boldmath $\alpha$}}}

\def\oper{{\mathchoice{\rm 1\mskip-4mu l}{\rm 1\mskip-4mu l}%
{\rm 1\mskip-4.5mu l}{\rm 1\mskip-5mu l}}}
\def\<{\langle}
\def\>{\rangle}
\newtheorem{theorem}{Theorem}

\newtheorem{Proposition}{Proposition}

\newtheorem{remark}{Remark}

\numberwithin{equation}{section}

\begin{document}
\title{\bf Geometry of quantum states: new construction of positive maps }
\author{Dariusz
Chru\'sci\'nski and Andrzej Kossakowski \\
Institute of Physics, Nicolaus Copernicus University,\\
Grudzi\c{a}dzka 5/7, 87--100 Toru\'n, Poland}

\maketitle

\begin{abstract}

We provide a new class of positive maps in matrix algebras. The
construction is based on the family of balls living in the space of
density matrices of $n$-level quantum system. This class generalizes
the celebrated Choi map and provide a wide family of entanglement
witnesses which define a basic tool for analyzing quantum
entanglement.

\end{abstract}

\maketitle

\section{Introduction}

One of the most important problems of quantum information theory
\cite{QIT} is the characterization of mixed states of composed
quantum systems. In particular it is of primary importance to test
whether a given quantum state exhibits quantum correlation, i.e.
whether it is separable or entangled. For low dimensional systems
there exists simple necessary and sufficient condition for
separability. The celebrated Peres-Horodecki criterium
\cite{Peres,PPT} states that a state of a bipartite system living in
$\mathbb{C}^2 \ot \mathbb{C}^2$ or $\mathbb{C}^2 \ot \mathbb{C}^3$
is separable iff its partial transpose is positive. Unfortunately,
for higher-dimensional systems there is no single universal
separability condition.

The most general approach to separability problem is based on the
following observation \cite{Horodeccy-PM}: a state $\rho$ of a
bipartite system living in $\mathcal{H}_A \ot \mathcal{H}_B$ is
separable iff $\mbox{Tr}(W\rho) \geq 0$ for any Hermitian operator
$W$ satisfying $\mbox{Tr}(W P_A \ot P_B)\geq 0$, where $P_A$ and
$P_B$ are projectors acting on $\mathcal{H}_A$ and $\mathcal{H}_B$,
respectively. Recall, that a Hermitian  operator $W \in
\mathcal{B}(\mathcal{H}_A \ot \mathcal{H}_B)$ is an entanglement
witness \cite{Horodeccy-PM,Terhal1} iff: i) it is not positively
defined, i.e. $W \ngeq 0$, and ii) $\mbox{Tr}(W\sigma) \geq 0$ for
all separable states $\sigma$. A bipartite state $\rho$ living in
$\mathcal{H}_A \ot \mathcal{H}_B$ is entangled iff  there exists an
entanglement witness $W$ detecting $\rho$, i.e. such that
$\mbox{Tr}(W\rho)<0$. Clearly, the construction of entanglement
witnesses is a hard task. It is easy to construct $W$ which is not
positive, i.e. has at leat one negative eigenvalue, but it is very
difficult to check that $\mbox{Tr}(W\sigma) \geq 0$ for all
separable states $\sigma$.

The separability problem may be equivalently formulated in terms
positive maps \cite{Horodeccy-PM}: a state $\rho$ is separable iff
$(\oper \ot \Lambda)\rho$ is positive for any positive map $\Lambda$
which sends positive operators on $\mathcal{H}_B$ into positive
operators on $\mathcal{H}_A$.  Positive maps play important role
both in physics and mathematics providing generalization of
$*$-homomorphism, Jordan homomorphism and conditional expectation.
Normalized positive maps define an affine mapping between sets of
states of $\mathbb{C}^*$-algebras. Unfortunately, in spite of the
considerable effort (see e.g.  \cite{Arveson}--\cite{atomic}), the
structure of positive maps (and hence also the set of entanglement
witnesses) is rather poorly understood.

In the present paper we construct a new class of positive maps using
the family of balls contained in the space of density matrices of
$n$-level quantum system.  Our construction generalizes a class of
maps introduced in \cite{Kossak1}.

The paper is organized as follows: in the next Section we introduce
a family of balls in the space of quantum states. We show that each
faithful state (i.e. strictly positive density operator) serves as a
center of the ball. In particular ball centered at the maximally
mixed state $\rho_0=\mathbb{I}/n$ possesses a maximal radius
$[n(n-1)]^{-1/2}$. Section \ref{S3} provides positive maps with
values in the corresponding ball. Composing with affine maps they
give rise to the wide class of positive maps discussed in Section
\ref{S4}. Finally, in Section \ref{S-Choi} we illustrate our
construction for $n=3$ and provide generalization of the celebrated
Choi map \cite{Choi}. A brief discussion is included in the last
section.

\section{A family of balls}

Let us consider the space of quantum states $\mathcal{S}_n$
corresponding to $n$-level quantum system, i.e. the space of density
operators living in the Hilbert space $\mathcal{H} = \mathbb{C}^n$.
It defines a convex subset of the linear space of Hermitian
operators
\begin{equation}\label{}
    H_n = \{ \, a \in M_n(\mathbb{C})\, |\, a^*=a\, \}\ ,
\end{equation}
where $M_n(\mathbb{C})$ denotes the space of $n \times n$ complex
matrices. Recall, that $H_n$ is a real Hilbert space equipped with
the scalar product $(a,b)= \mbox{tr}(ab)$ and the norm
$|\!|a|\!|^2=(a,a)$. Now, let $\widetilde{\rho} \in \mathcal{S}_n$
be a strictly positive density matrix, i.e. its spectral
decomposition has the following form
\begin{equation}\label{rho-tilde}
\widetilde{\rho} = \widetilde{\lambda}_1P_1 + \widetilde{\lambda}_2
P_2 + \ldots + \widetilde{\lambda}_n P_n \ ,
\end{equation}
where
\begin{equation}\label{>0}
    \widetilde{\lambda}_1 \geq \widetilde{\lambda}_2 \geq \ldots
    \geq \widetilde{\lambda}_n > 0 \ .
\end{equation}
A set of rank 1 projectors $\mathbf{P}=\{P_1,P_2,\ldots,P_n\}$
defines a simplex $\Sigma(\mathbf{P}) \subset \mathcal{S}_n$, and
the condition (\ref{>0}) implies that $\widetilde{\rho}$ belongs to
the interior of $\Sigma({\bf P})$. Note, that $\widetilde{\rho}$ may
be rewritten as follows
\begin{equation}\label{rho-tilde-new}
\widetilde{\rho} = {\lambda}_1P_1 + {\lambda}_2 P_2 + \ldots +
{\lambda}_{n-1} P_{n-1} + {\lambda_n}\frac{\mathbb{I}}{n} \ ,
\end{equation}
where
\begin{equation}\label{}
    \lambda_i = \widetilde{\lambda}_i - \widetilde{\lambda}_n \geq 0  \ ; \
    \ \ \ i=1,\ldots,n-1\ ,
\end{equation}
and
\begin{equation}\label{}
    \lambda_n = n \widetilde{\lambda}_n > 0 \ .
\end{equation}
 Let $\mbox{F}_i$ be a $(n-2)$-dimensional face of $\Sigma({\bf
P})$, i.e. a set
\begin{equation}\label{}
    \mbox{F}_i({\bf P}) = \Big\{\, \sum_{k=1}^n p_k P_k \subset \Sigma({\bf P}) \, \Big|\,
    p_i = 0 \, \Big\}\ ,
\end{equation}
and for any $a \in H_n$ and $r>0$ denote by $B_n(a,r)$ the following
ball
\begin{equation}\label{}
    B_n(a,r) = \{ x\in H_n \, | \, || a - x || \leq r \, \}\, \subset\, H_n\ .
\end{equation}

\begin{theorem}
For any $r \leq r_{\rm max} := \lambda_n/\sqrt{n(n-1)}$ one has
$B_n(\widetilde{\rho},r) \subset \Sigma({\bf P})$. Moreover, a
maximal ball $B_n(\widetilde{\rho},r_{\rm max})$ is tangent to the
face ${\rm F}_n({\bf P})$.
\end{theorem}

\noindent {\bf Remark.} In the special case when $\widetilde{\rho}
\equiv \rho_0 = \mathbb{I}/n$, one has $\lambda_n = 1$ and $r_{\rm
max}  = 1/\sqrt{n(n-1)}$ defines a ball $B_n(\rho_0,r_{\rm max})$
inscribed in each simplex ${\bf P}=\{ P_1,\ldots,P_n\}$
\cite{Kossak1}, that is,  this ball is tangent to each face ${\rm
F}_i({\bf P})$.

\vspace{.5cm}

\noindent {\bf Proof.} Take an arbitrary point $\rho_\sbalpha \in
{\rm F}_n({\bf P})$, i.e.
\begin{equation}\label{}
    \rho_\sbalpha = \alpha_1 P_1 + \ldots + \alpha_{n-1}P_{n-1}\ ,
\end{equation}
with $\alpha_i\geq 0$, and $\alpha_1 + \ldots + \alpha_{n-1}=1$. Let
us compute a distance between $\widetilde{\rho}$ and $\rho_\sbalpha$
\begin{equation}\label{}
    D(\!\balpha):= || \widetilde{\rho} - \rho_\sbalpha||^2 \ .
\end{equation}
One finds
\begin{equation}\label{D-a}
D(\!\balpha) = (\alpha_1 - \lambda_1)^2 + \ldots + (\alpha_{n-1} -
\lambda_{n-1})^2 - \frac{\lambda_n^2}{n}\ .
\end{equation}
To find a minimum of $D(\!\balpha)$ we treat
$\alpha_1,\ldots,\alpha_{n-2}$ as independent variables
($\alpha_{n-1}= 1 - \alpha_1 - \ldots - \alpha_{n-2}$). The
condition for a local extremum
\begin{equation}\label{}
    \frac{\partial D(\!\balpha)}{\partial \alpha_i} = 0 \ ; \ \ \
    \ i=1,\ldots,n-2\ ,
\end{equation}
gives rise to the following system of linear equations
\begin{equation}\label{}
    \sum_{j=1}^{n-2} {\bf A}_{ij} \, \alpha^*_j = \beta_i \ ; \ \ \ i=1,\ldots,n-2\ ,
\end{equation}
where the $(n-2)\times (n-2)$ matrix ${\bf A}$ reads as follows
\begin{equation}\label{}
    {\bf A}_{ij} = \left\{ \begin{array}{cl} n-2 &; \ \  i=j \\ 1 &; \ \
 i\neq j \end{array} \right. \ ,
\end{equation}
and
\begin{equation}\label{}
    \beta_i = 1 + \lambda_i - \lambda_{n-1} \ ; \ \ \ i=1,\ldots,n-2\
    .
\end{equation}
Finding the inverse matrix
\begin{equation}\label{}
    {\bf A}^{-1}_{ij} =  \left\{ \begin{array}{ll} \frac{n-2}{n-1} &; \ \  i=j \\ \frac{-1}{n-1}\, & ; \ \
    i\neq j \end{array} \right. \ ,
\end{equation}
one obtains for the solution
\begin{equation}\label{}
    \alpha^*_i = \lambda_i + \frac{\lambda_n}{n-1} \ ; \ \ \ i=1,\ldots,n-1\
    .
\end{equation}
Inserting $\balpha^* = (\alpha^*_1,\ldots,\alpha^*_{n-1})$ into
(\ref{D-a}) one finds
\begin{equation}\label{r-max}
    r^2_{\rm max} := D(\!\balpha^*) = \frac{\lambda_n^2}{n(n-1)}\ ,
\end{equation}
which ends the proof. \hfill $\Box$

\section{From balls to positive maps} \label{S3}

Let us consider the following linear map
\begin{equation}\label{}
    \varphi_\mu \ : \ M_n(\mathbb{C}) \ \longrightarrow\
    M_n(\mathbb{C})\ ,
\end{equation}
defined by
\begin{equation}\label{}
    \varphi_\mu(a) := \mu a + (1-\mu )\widetilde{\rho}\, \mbox{tr}\, a \ ,
\end{equation}
with a real parameter $\mu$. Note, that
\begin{equation}\label{}
\varphi_\mu(\widetilde{\rho}) = \widetilde{\rho}\ ,
\end{equation}
and
\begin{equation}\label{}
    {\rm tr}\, \varphi_\mu(a) = {\rm tr}\, a\ .
\end{equation}
It is clear that if $\mu\in [0,1]$ then $\varphi_\mu$ is a CP map
being a convex combination of two CP maps. Our aim is to prove the
following
\begin{theorem}
If $\mu$ satisfies
\begin{equation}\label{mu}
    |\mu| \leq \mu_{\rm max} \ ,
\end{equation}
where
\begin{equation}\label{mu-max}
    \mu_{\rm max} :=  \frac{r_{\rm max}}{\sqrt{1 + \lambda_1^2 + \ldots + \lambda_{n-1}^2 -
    \lambda_n^2/n }}\ ,
\end{equation}
and $\, r_{\rm max}$ is defined in (\ref{r-max}), then $\varphi_\mu$
is a positive map.
\end{theorem}

\vspace{.5cm}

\noindent {\bf Proof.} For any rank 1 projector $P$ one has
\begin{equation}\label{}
    \frac{\lambda_n}{n} \leq \mbox{tr}(\widetilde{\rho}P) \leq
    \frac{\lambda_n}{n} + (\lambda_1 + \ldots + \lambda_{n-1})\ .
\end{equation}
Now, for any $\rho \in \mathcal{S}_n$
\begin{equation}\label{}
    || \widetilde{\rho} - \rho || \leq \max_P\, ||\widetilde{\rho} -
    P||\ ,
\end{equation}
where the maximum is taken over all rank 1 projectors  $P\in
\mathcal{S}_n$. Now
\begin{equation}\label{}
    ||\widetilde{\rho} - P||^2 = ||\widetilde{\rho}||^2 + ||P||^2 -
    2 \mbox{tr}(\widetilde{\rho}P) \leq ||\widetilde{\rho}||^2 + 1 -
    2 \frac{\lambda_n}{n}\ .
\end{equation}
Moreover, one easily finds
\begin{eqnarray}
% \nonumber to remove numbering (before each equation)
||\widetilde{\rho}||^2    &=& \lambda_1^2 + \ldots \lambda_{n-1}^2 + \frac{\lambda_n^2}{n} +
2\frac{\lambda_n}{n}(\lambda_1 + \ldots + \lambda_{n-1}) \nonumber  \\
   &=& \lambda_1^2 + \ldots \lambda_{n-1}^2 + 2\frac{\lambda_n}{n} -
   \frac{\lambda_n^2}{n} \ ,
\end{eqnarray}
and hence one obtains the following bound for the distance between
$\widetilde{\rho}$ and $P$
\begin{equation}\label{}
    || \widetilde{\rho} - P ||^2 \leq  \lambda_1^2 + \ldots \lambda_{n-1}^2 + 1 -
   \frac{\lambda_n^2}{n} \ .
\end{equation}
Now, let us compute the corresponding distance between
$\widetilde{\rho}$ and $\varphi_\mu(\rho)$ for an arbitrary state
$\rho \in \mathcal{S}_n$. Since
\begin{equation}\label{}
    \widetilde{\rho} - \varphi_\mu(\rho)   = \mu (\widetilde{\rho} -
    \rho)\ ,
\end{equation}
one has
\begin{eqnarray}
% \nonumber to remove numbering (before each equation)
  \max_\rho || \widetilde{\rho} - \varphi_\mu(\rho) ||^2 &=& \mu^2 \max_\rho || \widetilde{\rho} - \rho ||^2
  \leq \mu^2 \left( 1 - \frac{\lambda_n^2}{n} + \lambda_1^2 + \ldots
  + \lambda_{n-1}^2 \right) \ .
\end{eqnarray}
Now, assume that
\begin{equation}\label{mu1}
\mu^2 \left( 1 - \frac{\lambda_n^2}{n} + \lambda_1^2 + \ldots
  + \lambda_{n-1}^2 \right) \leq r_{\rm max}^2 =
  \frac{\lambda_n^2}{n(n-1)}\ .
\end{equation}
It implies that for any $\rho \in \mathcal{S}_n$ an image
$\varphi_\mu(\rho) \in B_n(\widetilde{\rho},r_{\rm max})$ and hence
$\tilde{\varphi}$ is a positive map. Formula (\ref{mu1}) is
equivalent to (\ref{mu}) which ends the proof. \hfill $\Box$

\vspace{.5cm}

\noindent {\bf Remark.} In the special case when $\widetilde{\rho}
\equiv \rho_0 = \mathbb{I}/n$, one has $\lambda_n = 1$ and
\begin{equation}\label{}
    \mu_{\rm max} = \frac{1}{n-1}\ ,
\end{equation}
which reproduces the result of \cite{Kossak1}.

\begin{figure}[t] \label{FIG}
\begin{center}
\epsfig{figure=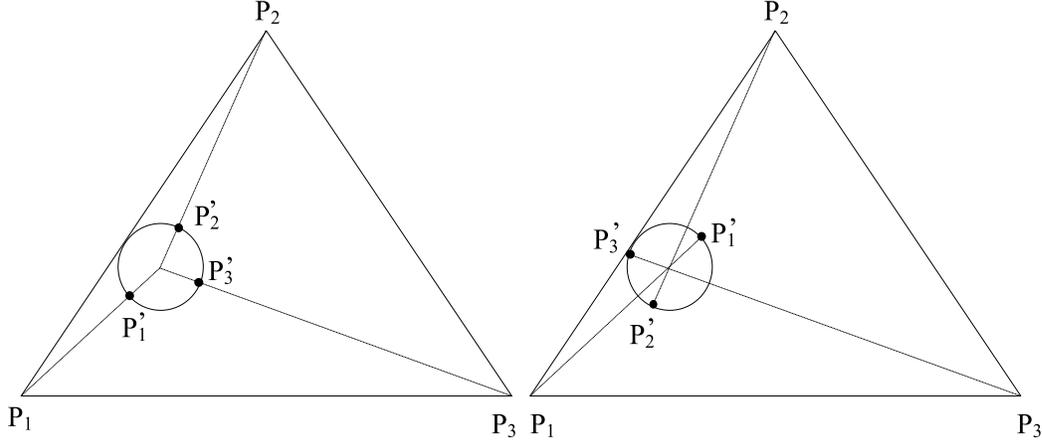,width=0.99\textwidth}
\end{center}
\caption{The action of $\varphi_\mu$ for $n=3$. It maps $P_k$ into
$P_k'$. On the left $\mu > 0$ and $\varphi_\mu$ is CP, on the right
$\mu <0$ and $\varphi_\mu$ is positive but not CP.}
\end{figure}

Figure 1 shows the action of $\varphi_{\mu}$ with $|\mu|=\mu_{\rm
max}$ for $n=3$, i.e. $\varphi_{\mu}(P_k) = P_k'$. The figure on the
left corresponds to ${\mu}> 0$ and the map $\varphi_{\mu}$ is
completely positive being a sum of two completely positive maps. The
figure on the right corresponds to ${\mu} <0$ and the the map
$\varphi_{\mu}$ is positive but not CP.

\section{Composing with affine maps} \label{S4}

Having define a map  $\varphi_\mu$ with a property that
$\varphi_\mu(\rho) \in B_n(\widetilde{\rho},r_{\rm max})$ for all
density operators $\rho \in \mathcal{S}_n$ let us observe that we
may compose it with an arbitrary affine  map  which maps a ball
$B_n(\widetilde{\rho},r_{\rm max})$ into itself, i.e. if
\begin{equation}\label{}
    \psi \ : \  B_n(\widetilde{\rho},r_{\rm max}) \ \longrightarrow\ B_n(\widetilde{\rho},r_{\rm
    max}) \ ,
\end{equation}
then $\psi \circ \varphi_\mu$ maps all density matrices from
$\mathcal{S}_n$ into $B_n(\widetilde{\rho},r_{\rm max})$.
 Denote by
${\rm Aff}_n$ a set of affine maps $({\rm T},{\bf t}) : \mathbb{R}^n
\longrightarrow\mathbb{R}^n$ which map the closed unit balls into
itself, i.e.
\begin{equation}\label{}
    ({\rm T},{\bf t}){\bf x} := {\rm T}{\bf x} + {\bf t} \ ,
\end{equation}
where ${\rm T} \in M_n(\mathbb{R})$ and ${\bf t} \in \mathbb{R}^n$
represents translation. Now, ${\rm Aff}_n$ being a compact convex
set it is entirely determined by its extremal elements.

\begin{Proposition}
The extremal elements ${\rm Extr}\, {\rm Aff}_n$ are defined by
\begin{equation}\label{}
    {\rm T} = R_1 \Lambda R_2\ , \ \ \ {\bf t} = R_1 {\bf c}\ ,
\end{equation}
where $R_1,R_2 \in O(n)$, $\Lambda$ is diagonal with eigenvalues
\[ \lambda_1 = \ldots = \lambda_{n-1} = \frac{\lambda_n}{\kappa} =
\sqrt{1-\delta^2(1-\kappa^2)} \ , \] with $0 \leq \kappa \leq 1\, $
and  $0 < \delta \leq 1\, $. Finally, ${\bf c} = (c_1,\ldots,c_n)$
reads as follows
\[  c_1 = \ldots = c_{n-1}=0\ , \ \ \ c_n = \delta(1-\kappa^2) \ .
\]
\end{Proposition}
For the proof see \cite{Gorini}. Note, that $({\rm T},r{\bf t})$
maps a ball with radius $r$ into itself provided $({\rm T},{\bf t})
\in {\rm Aff}_n$. Denote by ${\rm Aff}_n^0$ a subset of ${\rm
Aff}_n$ corresponding to $\kappa=0$. It is clear that
\begin{equation}\label{}
    {\rm Extr}\, {\rm Aff}^0_n = \{ \, ({\rm T},{\bf t}) \in {\rm Aff}_n\ : \ {\rm T}\in O(n)\ , {\bf t} =0\,
    \}\ .
\end{equation}
It is convenient to introduce an orthonormal basis in $H_n$: ${\bf
f} = (f_1,\ldots,f_{n^2-1})$ and $f_{n^2} = \mathbb{I}/\sqrt{n}$,
such that $(f_\alpha,f_\beta)=\delta_{\alpha\beta}$. It implies that
${\rm tr}\, f_\alpha = 0 $ for $\alpha = 1,\ldots,n^2-1$. Now, any
element $a \in H_n$ may be decomposed as follows
\begin{equation}\label{}
    a = \frac{\mathbb{I}}{n}\, {\rm tr}\, a + \< {\bf f},{\bf
    a}\>\ ,
\end{equation}
with ${\bf a}=(a_1,\ldots,a_{n^2-1}) \in \mathbb{R}^{n^2-1}\,$,
$a_\alpha = {\rm tr}(f_\alpha a)\,$, and $\< {\bf f},{\bf     a}\> =
\sum_{\alpha=1}^{n^2-1} f_\alpha a_\alpha $. In particular one has
\begin{equation}\label{x-tilde}
    \widetilde{\rho} = \frac{\mathbb{I}}{n} + \< {\bf f},\widetilde{{\bf
    x}}\>\ ,
\end{equation}
and
\begin{equation}\label{a-prim}
    a' := \varphi_\mu(a) = \frac{\mathbb{I}}{n}\, {\rm tr}\, a + \< {\bf f},{\bf
    a}'\>\ ,
\end{equation}
due to ${\rm tr}\, \varphi_\mu(a)={\rm tr}\, a$. Now, if $a' \in
B(\widetilde{\rho},r_{\rm max})$ we may shift ${\bf a}'$ by
`$-\widetilde{\bf x}\,{\rm tr}a$', apply an affine map $({\rm
T},r_{\rm max}{\bf t})$ and then shift back by `$\widetilde{\bf
x}\,{\rm tr}a$'. As a result one obtains again an element $a'' \in
B(\widetilde{\rho},r_{\rm max})$. Therefore, the main result of this
section may be summarized by the following
\begin{theorem}
For $|\mu|\leq \mu_{\rm max}\,$ every affine map $({\rm T},{\bf t})
\in {\rm Aff}_{n^2-1}$ induces a positive trace preserving map
\[   \varphi_\mu[{\rm T},{\bf t}] \ : \ M_n(\mathbb{C}) \ \longrightarrow\
M_n(\mathbb{C}) \ , \] defined by
\begin{equation}\label{BASIC}
    \varphi_\mu[{\rm T},{\bf t}](a)  = \widetilde{\rho}\, {\rm tr}\, a + \< {\bf f},({\rm T},r_{\rm max}{\bf t})({\bf
    a}' - \widetilde{\bf x}{\rm tr}a)\>\ ,
\end{equation}
where $\widetilde{\bf x}\,$ and ${\bf a}'$ are given by
(\ref{x-tilde}) and (\ref{a-prim}), respectively.
\end{theorem}

\begin{remark} {\em Actually, we have constructed  the action of $ \varphi_\mu[{\rm T},{\bf t}]$ only
for hermitian elements. However, due to the linearity one obviously
has
\begin{equation}\label{}
\varphi_\mu[{\rm T},{\bf t}](a) = \varphi_\mu[{\rm T},{\bf t}](a_1)
+ i\, \varphi_\mu[{\rm T},{\bf t}](a_2)\ ,
\end{equation}
where $a = a_1 + ia_2$ is an arbitrary element from
$M_n(\mathbb{C})$ with $a_1,a_2 \in H_n$. }
\end{remark}

\begin{remark} {\rm If $\widetilde{\rho}= \mathbb{I}/n$, then one
recovers a family of positive maps constructed in \cite{Kossak1}. }
\end{remark}

\section{Example: generalized Choi map}  \label{S-Choi}

Our basic formula (\ref{BASIC}) does depend upon an orthonormal
basis $f_\alpha$. Now, let $\{e_1,\ldots,e_n\}$ denote the
eigen-basis of $\widetilde{\rho}$, that is, $\widetilde{\rho} e_i =
\widetilde{\lambda}_i e_i$. Let us construct ${\bf f}
=(f_1,\ldots,f_{n^2-1})$ as the following generators of $SU(n)$
\[ (f_1,\ldots,f_{n^2-1}) = (d_\ell,u_{ij},v_{ij}) \ , \]
with $\ell = 1,\ldots,n-1\,$, and $1 \leq i < j \leq n\ $: $d_\ell$
generate Cartan subalgebra
\begin{equation}\label{}
    d_\ell = \frac{1}{\sqrt{\ell(\ell+1)}}\Big( \sum_{k=1}^\ell  e_{kk} -
    \ell e_{\ell+1,\ell+1} \Big)\ ,
\end{equation}
and
\begin{equation}\label{}
    u_{ij} = \frac{1}{\sqrt{2}}\, (e_{ij} + e_{ji})\ , \ \ \ \  v_{ij} = \frac{1}{\sqrt{2}\, i}\, (e_{ij} - e_{ji})\ ,
\end{equation}
where  $e_{ij} := |e_i\>\<e_j|$.

To illustrate our general scheme let us consider $n=3$ and take an
affine transformation from a set ${\rm Extr}\, {\rm Aff}^0_{8}$,
i.e. $({\rm T},{\bf t})$ with ${\rm T} \in O(8)$ and ${\bf t}=0$.
Let us introduce the following set of coordinates in $\mathbb{R}^8$:
\begin{equation}\label{}
    x_\ell = {\rm tr}(ad_\ell)\ , \ \ \ \ell = 1,2\ ,
\end{equation}
and
\begin{equation}\label{}
    x_{ij} = {\rm tr}(a u_{ij}) , \ \ \ \ y_{ij} =   {\rm tr}(a v_{ij})\
    , \ \ \ 1 \leq i < j \leq 3\ .
\end{equation}
Now, let ${\rm T}$ be a rotation from $O(8)$ given by
\begin{eqnarray}
% \nonumber to remove numbering (before each equation)
  x'_1 &=& x_1 \cos\alpha - x_2 \sin\alpha \nonumber\ , \\
  x'_2 &=& x_1 \sin\alpha + x_2 \cos\alpha \nonumber\ , \\
  x'_{ij} &=& - x_{ij} \ , \\
  y'_{ij} &=& - y_{ij}\ . \nonumber
\end{eqnarray}
In this parametrization the map
\[   \varphi_{\mu_{\rm max}}[\alpha] \ : \ M_3(\mathbb{C}) \ \longrightarrow\
M_3(\mathbb{C}) \ , \] has the following form
\begin{eqnarray}
% \nonumber to remove numbering (before each equation)
   \varphi_{\mu_{\rm max}}[\alpha](e_{ii})  &=& \sum_{j=1}^3 \Lambda_{ij} e_{jj} \ , \\
   \varphi_{\mu_{\rm max}}[\alpha](e_{ij}) &=& -\mu_{\rm max} e_{ij} \ , \ \ \ i\neq j \ ,
\end{eqnarray}
where
\begin{equation}\label{}
    \Lambda_{ij} = \mu_{\rm max} \Lambda^0_{ij} + \Lambda^1_{ij} \ ,
\end{equation}
with $\Lambda^0$ being a circulant matrix defined by
\begin{equation}\label{}
    \Lambda^0 = \left( \begin{array}{ccc} \eta_1 & \eta_2 & \eta_3
    \\ \eta_3 & \eta_1 & \eta_2 \\ \eta_2 & \eta_3 & \eta_1
    \end{array} \right) \ ,
\end{equation}
where the matrix elements $\eta_i$ depend upon the parameter
$\alpha$ in the following way
\begin{eqnarray}
% \nonumber to remove numbering (before each equation)
  \eta_1(\alpha) &=& \frac 23 \, \cos\alpha\nonumber \ , \\
  \eta_2(\alpha) &=& - \frac 13 ( \cos\alpha + \sqrt{3}\, \sin\alpha) \ , \\
  \eta_3(\alpha) &=& \frac 13\, ( - \cos\alpha + \sqrt{3}\, \sin\alpha) \nonumber \
  ,
\end{eqnarray}
and
\begin{equation}\label{}
    \Lambda^1 = \left( \begin{array}{ccc} \xi_1 & \xi_1 & \xi_1
    \\ \xi_2 & \xi_2 & \xi_2 \\ \xi_3 & \xi_3 & \xi_3
    \end{array} \right) \ ,
\end{equation}
with
\begin{eqnarray}
% \nonumber to remove numbering (before each equation)
  \xi_1 &=& \lambda_1 + \frac{\lambda_3}{3} - \mu_{\rm max} \Big[ \lambda_1 \eta_1(\alpha) + \lambda_2 \eta_2(\alpha) \Big] \nonumber \ , \\
\xi_2 &=& \lambda_2 + \frac{\lambda_3}{3} - \mu_{\rm max} \Big[ \lambda_1 \eta_3(\alpha) + \lambda_2 \eta_1(\alpha) \Big]  \ , \\
\xi_3 &=& \ \ \ \ \ \ \, \frac{\lambda_3}{3} - \mu_{\rm max} \Big[
\lambda_1 \eta_2(\alpha) + \lambda_2 \eta_3(\alpha) \Big] \nonumber
\ .
\end{eqnarray}
 Note that
\begin{equation}\label{eee}
\eta_1(\alpha) + \eta_2(\alpha) + \eta_2(\alpha) = 0 \ ,
\end{equation}
and
\begin{equation}\label{}
    \xi_1 + \xi_2 + \xi_3 = \lambda_1 + \lambda_2 + \lambda_3 = 1\ .
\end{equation}
The matrix $\Lambda^0$ is {\em universal}, i.e. it does not depend
upon the invariant state $\widetilde{\rho}$.

\begin{remark}
{\em  If $\widetilde{\rho} = \mathbb{I}/3$, then
\begin{equation}\label{}
    \xi_1 = \xi_2 = \xi_3 = \frac 13\ ,
\end{equation}
and the matrix $\Lambda$ is circulant and stochastic (hence doubly
stochastic). For $\widetilde{\rho} \neq \mathbb{I}/3$, it is no
longer circulant but $\Lambda^T$ is stochastic. }
\end{remark}

\begin{remark}
{\em The map $\varphi_{\mu_{\rm max}}[\alpha=\pi/3]$ reduces for
$\widetilde{\rho} = \mathbb{I}/3$ to  the celebrated Choi map
\cite{Choi} defined by
\begin{eqnarray}
% \nonumber to remove numbering (before each equation)
   \varphi_{\rm Choi}(e_{ii})  &=& \sum_{j=1}^3 \Lambda^{\rm Choi}_{ij} e_{jj} \ , \\
   \varphi_{\rm Choi}(e_{ij}) &=& -\frac 12\, e_{ij} \ , \ \ \ i\neq j \ ,
\end{eqnarray}
where the doubly stochastic matrix $\Lambda^{\rm Choi}$ is defined
by
\begin{equation}\label{}
    \Lambda^{\rm Choi} = \frac 12 \left( \begin{array}{ccc} 1 & 1 &
    0    \\ 0 & 1 & 1 \\ 1 & 0 & 1
    \end{array} \right) \ .
\end{equation}
}
\end{remark}

\vspace{.3cm}

 Finally, let us note that the corresponding
entanglement witness \[ W[\alpha] = 3({\rm id} \ot \varphi_{\mu_{\rm
max}}[\alpha])P^+_3\ ,
\]
 where $P^+_3$ denotes the maximally entangled state in $\mathbb{C}^3 \ot
\mathbb{C}^3$, reads as follows
\begin{equation}\label{}
 \hspace*{-.1cm}
  W[\alpha]\, =\, \mu_{\rm max}\, \left( \begin{array}{ccc|ccc|ccc}
    a_1 & \cdot & \cdot & \cdot & -1 & \cdot & \cdot & \cdot & -1 \\
    \cdot& b_{1} & \cdot & \cdot & \cdot& \cdot & \cdot & \cdot & \cdot  \\
    \cdot& \cdot & c_{1} & \cdot & \cdot & \cdot & \cdot & \cdot &\cdot   \\ \hline
    \cdot & \cdot & \cdot & c_{2} & \cdot & \cdot & \cdot & \cdot & \cdot \\
    -1 & \cdot & \cdot & \cdot & a_{2} & \cdot & \cdot & \cdot & -1  \\
    \cdot& \cdot & \cdot & \cdot & \cdot & b_{2} & \cdot & \cdot & \cdot  \\ \hline
    \cdot & \cdot & \cdot & \cdot& \cdot & \cdot & b_{3} & \cdot & \cdot \\
    \cdot& \cdot & \cdot & \cdot & \cdot& \cdot & \cdot & c_{3} & \cdot  \\
    -1 & \cdot& \cdot & \cdot & -1 & \cdot& \cdot & \cdot & a_{3}
     \end{array} \right)\ ,
\end{equation}
where the $\alpha$-dependent coefficients are given by
\begin{eqnarray}
% \nonumber to remove numbering (before each equation)
  a_i = \frac{\eta_1(\alpha) + \xi_i}{\mu_{\rm max}} \ , \ \ \ \ \
  b_i = \frac{\eta_2(\alpha) + \xi_i}{\mu_{\rm max}}\ , \ \ \ \ \
  c_i = \frac{\eta_3(\alpha) + \xi_i}{\mu_{\rm max}}\ .
\end{eqnarray}
It is clear that $a_i,b_i,c_i \geq 0$ and
\begin{equation}\label{}
    a_i + b_{i+1} + c_{i+2} = \frac 1\mu_{\rm max}\ ,
\end{equation}
for  $i=1,2,3$ (mod 3). The above class of entanglement witnesses
belongs to a class of bipartite operators studied in
\cite{PPT-nasza}. Note, that $W[\alpha]$ defines true entanglement
witness iff it is not positive, i.e. possesses at least one negative
eigenvalue, that is, the following $3 \times 3 $ matrix
\begin{equation*}\label{}
\left( \begin{array}{ccc}  a_1 & -1 &  -1 \\ -1 & a_2 & -1 \\ -1 &
-1 & a_3  \end{array} \right)
\end{equation*}
is not positive. It is easy to see that if $\widetilde{\rho} =
\mathbb{I}/3$, then $W[\alpha]$ is never positive. However, it is no
longer true for the general invariant state $\widetilde{\rho}$.

\section{Conclusions}

We introduced a new class of positive maps in the matrix algebra
$M_n(\mathbb{C})$ using a family of balls in the space of density
operators of $n$-level quantum system. Each map has an invariant
state $\widetilde{\rho}$ which defines the center of the ball. If
$\widetilde{\rho} = \mathbb{I}/n$, i.e. the map is unital, our
construction generalizes the family of positive maps introduced in
\cite{Kossak1}. In particular for $n=3$ it generalizes the
celebrated Choi map \cite{Choi}. As is well know positive maps which
are not completely positive provide a basic tool to study quantum
entanglement. Therefore our method provides new class of
entanglement witnesses.

Presented construction guarantees positivity but says nothing about
indecomposability and/or optimality \cite{optimal}. Both
indecomposable and optimal positive maps are crucial in detecting
and classifying quantum entanglement. Therefore, the analysis of
positive maps based on the family of balls deserves further study.

 We stress that the structure off balls discussed in this paper may be easily
introduced for the composed $n \ot n$ system. In this case it
generalizes well known ball of separable states centered at
$\mathbb{I}/n^2$ \cite{Barnum}. It would be interesting to
investigate properties of quantum states belonging to other (not
necessary central) balls.

\section*{Acknowledgement} This work was partially supported by the
Polish Ministry of Science and Higher Education Grant No
3004/B/H03/2007/33 and by the Polish Research Network  {\it
Laboratory of Physical Foundations of Information Processing}.

\end{document}